\documentclass[%
 reprint,
 amsmath,amssymb,
 aps,
 pra,
]{revtex4-1}

\usepackage{graphicx}
\usepackage{dcolumn}
\usepackage{bm}
\usepackage[colorlinks=true]{hyperref}

\begin{document}

\title{Non-Markovian decoherence of a two-level system in a Lorentzian bosonic reservoir and a stochastic environment with finite correlation time  }

\author{V.\:A.\:Mikhailov}
\author{N.\:V.\:Troshkin}
\affiliation{%
 Department of Physics, Samara National Research University, 443086 Samara, Russia
}%

\date{\today}

\begin{abstract}
In this paper we investigate non-Markovian evolution of a two-level system (qubit) in a bosonic bath under influence of an external classical fluctuating environment. The interaction with the bath has the Lorentzian spectral density, and the fluctuating environment (stochastic field) is represented by a set of Ornstein-Uhlenbeck processes. Each of the subenvironments of the composite environment is able to induce non-Markovian dynamics of the two-level system. By means of the numerically exact method of hierarchical equations of motion, we study dependence of the steady states of the two-level system, the reduced density matrix evolution and the equilibrium emission spectrums on frequency cutoffs and coupling strengths of the subenvironments. Additionally we investigate the impact of the rotation-wave approximation (RWA) used for the interaction with the bath on accuracy of the results.
   
\end{abstract}

\maketitle


\section{\label{sec:intro}Introduction}

Almost every quantum system interacts with its surroundings in a way that makes the system nearly impossible to be fully isolated. The interaction gives rise to processes of decoherence and appears to be harmful in some circumstances, e.g. in quantum information processing \cite{10.1088/0953-8984/28/21/213001, PhysRevA.99.022107, 10.1140/epjst/e2018-800081-0, 10.1038/s41598-018-19977-9}, and to be a valuable resource in others \cite{Ban2020, Moreira2020, Maier2019}. In some cases it is possible to capture evolution of the quantum system by means of a master equation of Lindblad form \cite{breuer2002theory, Rivas_2014, doi:10.1007/BF01608499}. Systems of that type are called Markovian, and their evolution has the form of a quantum Markov process. Markovian systems are common in quantum optics, where a quantum system is often weakly coupled to an environment characterized by negligibly small correlation time. 

Markovian systems assume their environments memoryless, however, an environment always changes under influence of a quantum system, which causes the memory effects to appear. Systems characterized by memoryful environments belong to the much wider category of non-Markovian systems \cite{Rivas_2014, RevModPhys.89.015001}. Due to  increased experimental and computational capabilities, non-Markovian systems are of great interest today. Among them there are such well-known systems as quantum dots \cite{10.1088/0022-3727/48/36/363001, 10.1088/1361-648x/aafd6a, PhysRevLett.112.147404, Bera20102260}, micromechanical resonators \cite{Aspelmeyer20141391, 10.1038/ncomms8606}, superconducting qubits \cite{Andersson20191123, 10.1038/s41467-018-03312-x, Yu2019}, and many others. Non-Markovian effects are ubiquitous in physics, chemistry, and biology and for systems interacting with either bosonic or fermionic reservoirs, like photosynthetic systems \cite{Pfalzgraff2019, Hwang-Fu201546, 10.1038/nphys2515, doi:10.1146/annurev-physchem-040215-112252, doi:10.1146/annurev-physchem-040215-112103}, molecular aggregates \cite{doi:10.1063/1.4846275}, molecular magnets \cite{Coish20092203}, and solar cells \cite{barford2013electronic}. Recently non-Markovian environments started to gain attention in quantum information processing \cite{PhysRevA.94.052115,10.1038/srep05720,10.1209/0295-5075/107/54006}, where attempts are made to utilize the backflow of information from the environment.   

Accurate description of non-Markovian systems is a more complicated problem in comparison with description of Markovian systems. There are a plenty of methods known, but none of them are generally applicable, i.e. each one has its strong and weak sides, and the number of systems it can describe efficiently is often limited. Analytical methods are represented mostly by perturbative expansions for some special parameter regimes, e.g. effective weak coupling theories or the projection operator techniques \cite{PhysRevB.84.081305,doi:10.1063/1.3247899}. Numerically exact methods include the ones based on enlarging of the system state space, e.g. by extending the system space by the most relevant environmental modes \cite{PhysRevA.55.2290, PhysRevA.101.052108, PhysRevLett.120.030402}, the ones utilizing tensor network approximations in propagation of influence functionals \cite{doi:10.1063/1.469508,doi:10.1063/1.469509,10.1038/s41467-018-05617-3} and in mappings on effective 1D fermionic and bosonic chains \cite{PhysRevLett.105.050404, PhysRevLett.123.090402, PhysRevB.101.155134}, and etc. One of the most well-established numerically exact methods is the method of hierarchical equations of motion (HEOM) \cite{doi:10.1143/JPSJ.75.082001, 10.1143/JPSJ.58.101,doi:10.1143/JPSJ.75.082001,doi:10.1063/1.4890441}. HEOMs utilize infinite systems of recurrent differential equations to encode the memory kernel of system-environment interaction and are able to handle a great variety of environmental spectral densities.    

Switching from the Markovian approximation to a full non-Markovian description reveals many interesting phenomena. The most common one is the emergence of oscillations \cite{PhysRevA.89.012107, doi:10.1063/1.4939733, PhysRevA.99.052102}. Non-Markovianity is known to be able to affect steady states (equilibrium states) of a system, e.g. it causes non-canonical steady states to appear \cite{PhysRevA.90.032114}, and also it affects the correlation functions. Actually, if non-Markovianity of a quantum process is sufficiently high, the quantum regression theorem (QRT) stops giving reliable correlation functions \cite{PhysRevA.99.012303, PhysRevA.92.062306, PhysRevA.93.022119}, which leads, for example, to significant differences between the predicted and the actual emission spectrums. It is a common practice to use the rotating wave approximation (RWA) for interaction with Markovian environments. Often the RWA is still used when the evolution become non-Markovian and can cause problems if non-Markovianity of the evolution is sufficiently high \cite{10.1088/1751-8113/43/40/405304, RevModPhys.89.015001, PhysRevA.88.052111}. For example, wrongly used RWA may lead to incorrect shifts of the system frequencies \cite{PhysRevA.94.012110} or cutoff all non-Markovianity \cite{PhysRevA.88.052111}. 

When an environment is composite, i.e. consists of several subenvironments, it is possible to utilize one of the subenvironments to control decoherence of a quantum system. The case of a stochastic subenvironment as a control tool is rather popular in literature \cite{PhysRevA.97.012104, PhysRevA.91.022109,10.1038/srep02746,PhysRevE.69.051110,PhysRevA.91.013838, Semin_2020} and has similarities with dynamical decoupling schemes that alter the environment spectral density via filtering functions realized in sequences of laser impulses \cite{10.1088/0953-4075/44/15/154002}.

In the paper we investigate non-Markovian evolution of a two-level system (TSL) in a bosonic bath under influence of an additional external fluctuating environment. The bath spectral density function is chosen to be Lorentzian \cite{breuer2002theory, PhysRevA.95.042132, PhysRevA.98.012110}. The Lorentzian spectral density is suitable, for example, for interaction between a Jaynes-Cummings cell and a zero-temperature bosonic bath \cite{PhysRevA.81.062124}. The stochastic environment is chosen to be represented by a set of Ornstein-Uhlenbeck random processes. Following \cite{mikhailov2020nonmarkovian}, we derive a HEOM capable of handling both RWA and non-RWA couplings with the bath equally accurate. Assuming the bath and the fluctuating environment to be independent, we analyze steady states of the TLS, evolution of the reduced density matrix, and equilibrium emission spectrums. We investigate the dependence on frequency cutoffs and coupling strengths of the subenvironments spectral densities in both RWA and non-RWA cases and provide a comparison with the ones obtained in the Markovian approximation \cite{mikhailov2016master}.  

The paper is organized as follows. In Sec.~\ref{sec:model-hamiltonian} we introduce the model, in Sec.~\ref{sec:heom} we present the hierarchical equations of motion. Next, we study the TLS evoution numerically. In Sec.~\ref{sec:steady-states} we study steady states of the TLS, in Sec.~\ref{sec:dm-evol} we investigate evolution of the reduced density matrix, and in Sec.~\ref{sec:spectrum} we investigate the emission spectrums. Finally, we draw conclusions in Sec.~\ref{sec:conclusions}.

\section{\label{sec:model-hamiltonian}Model}
The full Hamiltonian for the system can be written as
\begin{equation}
\label{HAFB}
\hat{H}=\hat{H}_A+\sum_{k=1}^{\infty}\hbar\omega_{k}\hat{b}_k^{+}\hat{b}_k+\hat{H}_{\text{IB}}+\hat{H}_{\text{IF}},
\end{equation}
where $\hat{H}_A=\hbar\omega_0\hat{\sigma}_{+}\hat{\sigma}_{-}$ is the Hamiltonian for the free TLS, $\omega_0$ is the TLS frequency, $\hat{\sigma}_{+}$ and $\hat{\sigma}_{-}$ are the rising and the lowering operators of the TLS, respectively; $\hat{b}_k^+$ and $\hat{b}_k$ form a set of creation and annihilation operators describing the modes of the bosonic bath; $\hat{H}_{IF}$ is the Hamiltonian for the interaction of the TLS with the fluctuating environment (stochastic field) and $\hat{H}_{\text{IB}}$ describes the interaction between the TLS and the bath.

The Hamiltonian for interaction with the stochastic field is defined by the next expression 
\begin{equation}
\label{HIF}
\hat{H}_{\text{IF}}=\hbar\Omega(t)\hat{\sigma}_{+}\hat{\sigma}_{-}+\hbar[\xi(t)\hat{\sigma}_{+}+\bar{\xi}(t)\hat{\sigma}_{-}],
\end{equation}
where $\Omega(t)$, $\xi(t)$ are random functions, $\bar{\xi}(t)$ is the complex conjugation of $\xi(t)$. The interaction gives rise to two decoherence channels, a dephasing channel and a relaxation channel, originating from the first and the second terms in (\ref{HIF}), respectively. 

The random function $\Omega(t)$ is a real random process and $\xi(t)$ is a complex random process. We assume that $\Omega(t)$ and $\xi(t)$ are Markov processes of Ornstein-Uhlenbeck (OU) type \cite{fpeRisken1996}, and consider the real and imaginary parts of $\xi(t)$ as two independent real OU processes $\xi_1(t)$ and $\xi_2(t)$, respectively. Correlation functions of the random processes have the same form
\begin{equation}
\label{field-corr-func}
\langle\nu(t)\nu(t')\rangle = \frac{\Delta_{\nu}^2}{\gamma_{\nu}}e^{-\gamma_{\nu}|t-t'|},
\end{equation} 
where $\nu \in\{\Omega, \xi_1, \xi_2\}$, $\Delta_\nu$ is the standard deviation of the OU process and defines the coupling strength with the stochastic field, $1/\gamma_\nu$ is the correlation time of the OU process and $\gamma_\nu$ has the physical meaning of the cut-off frequency of the environment.

The Hamiltonian $\hat{H}_{\text{IB}}$ is used in two forms, the form corresponding to the full electric-dipole interaction (non-RWA) and the form representing the interaction in the rotating wave approximation (RWA). Introducing a new auxiliary TLS operator $\hat{a}$, we can combine both forms of $\hat{H}_{\text{IB}}$ in one expression
\begin{equation}
\label{HIB-both-in-one}
\hat{H}_{\text{IB}}=\sum _{k=1}^{\infty}(g_k\hat{a}\hat{b}_k^{+}+\bar{g}_k\hat{a}^+\hat{b}_k),
\end{equation} 
where $g_k$ are the TLS-bath coupling constants, $\hat{a}^+$ is the adjoint of $\hat{a}$. If $\hat{a}=\hat{\sigma}_++\hat{\sigma}_-$, (\ref{HIB-both-in-one}) describes the full interaction, for $\hat{a}=\hat{\sigma}_-$ it corresponds to the RWA interaction. 

For path integral methods it is more naturally to define an interaction with environment in the continuous form via the spectral density function, instead of utilizing the coupling constants $g_k$ directly. In the paper we consider the Lorentzian spectral density \cite{breuer2002theory, PhysRevA.95.042132, PhysRevA.81.062124, PhysRevA.98.012110}
\begin{equation}
\label{lorentzian-sd}
J_L(\omega)=\frac{1}{\pi}\frac{\hbar^2\Delta_B^2\gamma_B^2}{\left(\omega-\omega_0\right)^2+\gamma_B^2},
\end{equation} 
where \(\Delta_B\) defines the coupling strength, \(\gamma_B\) is the bath spectral width, also called the environment cutoff frequency. For the noise induced by the bath, $1/\gamma_B$ represents the correlation time. The parameters have close relation to the corespondent parameters of the stochastic field $\Delta_\nu$ and $\gamma_\nu$ (\ref{field-corr-func}) and generally have the same physical meaning in terms of impact on the TLS dynamics.

\section{\label{sec:heom}Hierarchical equations of motion}

Let us suppose that before the initial moment of time the TLS does not interact with the bath and the stochastic field and both of the subenvironments are at equilibrium. Then the total density matrix at $t=t_0$ has the factorized form
\begin{equation}
\label{factorized-initial-conditions}
\hat{\rho}_{\text{tot}}\left(\Omega,\xi,\bar{\xi},t_0\right)=P_{\text{eq}}(\Omega,\xi_1,\xi_2)\hat{\rho}^{(A)}(t_0)\otimes\hat{\rho}_{\text{eq}}^{(B)}\left(t_0\right),
\end{equation}
where $P_{\text{eq}}(\Omega,\xi_1,\xi_2)$ is the factorizable Gaussian equilibrium distribution function of the stochastic field, $\hat{\rho}^{(A)}(t_0)$ denotes the initial density matrix of the TLS, and $\hat{\rho}_{\text{eq}}^{(B)}(t_0)$ is the equilibrium bath density matrix at zero temperature.

For the factorized initial conditions (\ref{factorized-initial-conditions}) and the Lorentzian spectral density (\ref{lorentzian-sd}), we can obtain the HEOM by the steps presented in \cite{mikhailov2020nonmarkovian}, where we have to replace the high-temperature Drude spectral density with the Lorentzian spectral density and to take the limit $\beta\to\infty$ for the bath subenvironment. The HEOM expression has the same form and can be written as
\begin{align}
\label{atom-bath-field-heom}
\frac{\partial }{\partial t}\hat{\rho}_{\bm{m}}^{(A)}(t)=&-\frac{i}{\hbar} \hat{H}_A^{\text{x}}\hat{\rho}_{\bm{m}}^{(A)}(t)
\notag\\
&+\sum_{k\in\{\text{field}\}}[m_k \alpha_k^{(F)}\hat{\rho}_{\bm{m}}^{(A)}(t)+\hat{\Phi}_{F,k}^{(0)}\hat{\rho}_{\bm{m|_{k+1}}}^{(A)}(t)
\notag\\
&+m_k\hat{\Phi}_{F,k}^{(1)}\hat{\rho}_{\bm{m|_{k-1}}}^{(A)}(t)]
\notag\\
&+\sum_{k\in\{\text{bath}\}}[m_k \alpha_{k}^{(B)}\hat{\rho}_{\bm{m}}^{(A)}(t)+ \hat{\Phi}_{B,k}^{(0)}\hat{\rho}_{\bm{m|_{k+1}}}^{(A)}(t)
\notag\\
&+m_k\hat{\Phi}_{B,k}^{(1)}\hat{\rho}_{\bm{m|_{k-1}}}^{(A)}(t)],
\end{align}
where $\bm{m}$ combines the field and the bath indexes into one composite index. We assume that the first three components of $\bm{m}$ index the recursion relations for the stochastic field in the order $\{\Omega, \xi_1, \xi_2\}$. The last two components index the bath recursion relations. The special notation $\bm{m|_{k+1}}$ is used for the index $\bm{m}$ with the $k$-th component increased by $1$
\begin{equation}
\begin{split}
\bm{m}&=(m_1,m_2,\dotsc),
\\
\bm{m|_{k+1}}&=(m_1,m_2,\dotsc,m_k+1,\dotsc),
\end{split}
\end{equation}
$\hat{\rho}_{\bm{m}}^{(A)}(t)$ denotes the $\bm{m}$-th auxiliary density matrix, $\hat{H}_A^{\text{x}}$ is the commutator superoperator for the free TLS, $\hat{H}_A^{\text{x}} \hat{\rho} = \hat{H}_A \hat{\rho} - \hat{\rho} \hat{H}_A$. The actual TLS density matrix starts the recursion and has index $\bm{m}=\bm{0}$, $\hat{\rho}^{(A)}(t)=\hat{\rho}_{\bm{0}}^{(A)}(t)$. The constants \(\alpha_k^{(F)}\) and the operators \(\hat{\Phi}_{F,k}^{(0)}\) and \(\hat{\Phi}_{F,k}^{(1)}\) belong to the stochastic field part of the recursion relations and can be expressed via parameters of the stochastic field and the field coupling operators
\begin{equation}
\begin{split}
\alpha_k^{(F)}=&-\gamma_{\nu_k},
\\
\hat{\Phi}_{F,k}^{(0)}=&-\Delta_{\nu_k}(i/\hbar)\hat{V}_{F,\nu_k}^{\text{x}},
\\
\hat{\Phi}_{F,k}^{(1)}=&-\Delta_{\nu_k}(i/\hbar)\hat{V}_{F,\nu_k}^{\text{x}},
\end{split}
\end{equation}
where $\nu_k$ denotes the k-th element in $\{\Omega,\xi_1,\xi_2\}$, e.g. $\nu_1 = \Omega$, $\hat{V}_{F,\nu_k}^{\text{x}}$ are commutator superoperators, i.e. $\hat{V}_{F,\nu_k}^{\text{x}}\hat{\rho}=\hat{V}_{F,\nu_k}\hat{\rho }-\hat{\rho }\hat{V}_{F,\nu_k}$, for the corresponding operators on the TLS subspace
\begin{equation}
\begin{split}
\hat{V}_{F,\Omega}&=\hbar \hat{\sigma}_{+}\hat{\sigma}_{-},
\\
\hat{V}_{F,\xi_1}&=\hbar (\hat{\sigma}_{+}+\hat{\sigma}_{-}),
\\
\hat{V}_{F,\xi_2}&=i \hbar (\hat{\sigma}_{+}-\hat{\sigma}_{-}).
\end{split}
\end{equation}
The remaining coefficients $\alpha_k^{(B)}$, $\hat{\Phi}_{B,k}^{(0)}$, and $\hat{\Phi}_{B,k}^{(1)}$ define the bath part of the recurrence relations and depend on the bath spectral density. If the spectral density is Lorentzian (\ref{lorentzian-sd}), we have the next expressions for the coefficients
\begin{equation}
\begin{split}
\label{lorentz-sd-heom-coeffs}
\alpha_{k_1}^{(B)}&=-(\gamma_B - i \omega_0),
\\
\alpha_{k_2}^{(B)}&=-(\gamma_B+ i \omega_0),
\\
\hat{\Phi}_{B,k_1}^{(0)}&=-\lambda_B\hat{c}_1^{\text{x}},
\\
\hat{\Phi}_{B,k_2}^{(0)}&=-\lambda_B\hat{c}_2^{\text{x}},
\\
\hat{\Phi}_{B,k_1}^{(1)}&=-(\Delta_B^2/2)\hat{c}_2^{\text{R}},
\\
\hat{\Phi}_{B,k_2}^{(1)}&=(\Delta_B^2/2)\hat{c}_1,
\end{split}
\end{equation}
where \(\hat{c}_1=\hat{a}\) and \(\hat{c}_2=\hat{a}^+\), the superscript ``$\text{x}$" denotes the commutator superoperator defined earlier, and the superscript ``$\text{R}$" denotes the superoperator for action from the right, i.e. \(\hat{c}_2^R\hat{\rho} = \hat{\rho}\hat{c}_2\).

From (\ref{lorentz-sd-heom-coeffs}) it follows that the bath part of the HEOM cannot be transformed into the one-indexed form, in contrast with the stochastic field part \cite{10.1143/JPSJ.58.101} and the case of non-RWA interaction with the high-temperature Drude bath \cite{mikhailov2020nonmarkovian}, because there are two unequal constants $\alpha_{k_1}^{(B)}$ and $\alpha_{k_2}^{(B)}$. While the infinite-temperature Drude bath appears to be quite similar to the stochastic field as an environment for the TLS \cite{doi:10.1143/JPSJ.75.082001}, we expect that the zero-temperature Lorentzian bath and the stochastic field act on the TLS in qualitatively different ways.

\section{\label{sec:steady-states}Steady states}
\begin{figure*}
\includegraphics[width=\linewidth]{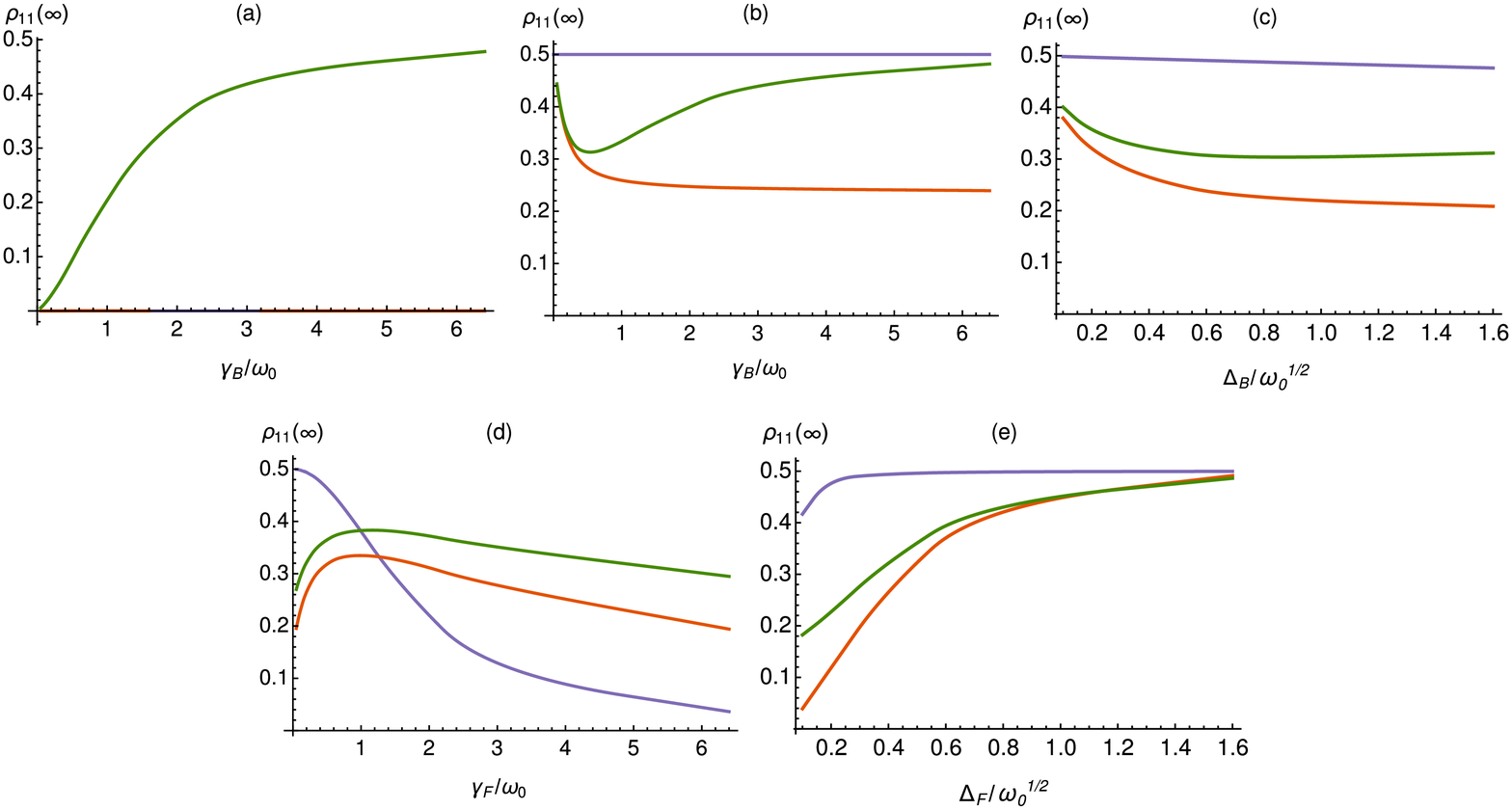}
\caption{
\label{fig:steady-states}
(a) The excited state population of the TLS in the steady states for interaction with the Lorentzian bath only (the field is off) as a function of the bath frequency cutoff $\gamma_B$, $\Delta_B/\omega_0^{1/2}=0.4$. Green denotes the non-RWA coupling with the bath. The RWA and the Markovian approximation curves coincide with the horizontal axis.
	(b, c) The excited state population of the TLS in the steady states for simultaneous interaction with the stochastic field and the Lorentzian bath as a function of (b) the bath frequency cutoff $\gamma_B$, $\Delta_B/\omega_0^{1/2}=0.4$, and (c) the bath coupling strength $\Delta_B$, $\gamma_B/\omega_0=0.8$. Orange and green denote the RWA and the non-RWA couplings, respectively, purple - the Markovian approximation. The stochastic field is characterized by $\gamma_\nu/\omega_0=0.2$ and $\Delta_\nu^2/\omega_0=0.4$.
	(d, e) The excited state population of the TLS in the steady states for simultaneous interaction with the stochastic field and the Lorentzian bath as a function of (d) the field frequency cutoff $\gamma_\nu=\gamma_F$, $\Delta_\nu/\omega_0^{1/2}=0.4$, and (e) the field coupling strength $\Delta_\nu=\Delta_F$, $\gamma_\nu/\omega_0=0.2$. Orange and green denote the RWA and non-RWA couplings respectively, purple - the Markovian approximation. The bath is characterized by $\gamma_B/\omega_0=0.8$, $\Delta_B/\omega_0^{1/2}=0.4$.
}
\end{figure*}

The steady states of the TLS reachable from the selected initial state (\ref{factorized-initial-conditions}) can be obtained by propagating the state forward in time until the reduced density matrix stops changing. Typical response of the steady states on changes of the environment parameters is presented in Fig.~\ref{fig:steady-states}. 

For a wide range of frequency cutoffs and coupling strengths tested, the stochastic field acting alone brings the TLS to the steady state where both the excited and the ground states are equally possible. Qualitatively different picture is observed for the TLS in the Lorentzian bath. Here, if one of the approximations is used, either the RWA or the Markovian, the TLS equilibrates in its ground state. Otherwise, in case of the non-RWA interaction with the bath, the probability of the excited state is above zero and gradually rises with the frequency cutoff, starting near zero and then approaching $1/2$ from below (Fig.~\ref{fig:steady-states}(a)). Also in the non-RWA case the stationary state exhibits a weak dependence on the coupling strength, the excited state probability rises from zero to $1/2$ from below, but much slower, then it is in the frequency cut-off case. Thus, the RWA, same as the Markovian approximation that intrinsically utilizes the RWA, alters the stationary states behavior drastically. 

When the TLS interacts with both subenvironments (Fig.~\ref{fig:steady-states}(b)-\ref{fig:steady-states}(e)), the difference between the stationary states for each of the subenvironments becomes visible. Because the difference between the steady states may be big, e.g. for low bath frequency cutoffs in the non-RWA case or for any parameters in the RWA case, the impact of the stochastic field can be significant.

If we fix parameters of the stochastic field and start increasing the bath frequency cutoff $\gamma_B$ (Fig.~\ref{fig:steady-states}(b)) or the coupling strength with the bath $\Delta_B$ (Fig.~\ref{fig:steady-states}(c)) starting from zero, the bath contribution in a steady state grows and the excited state population in the steady state decreases, because the steady states for decoherence in the bath always lie lower. Because in the RWA case the steady states are all completely unexcited (Fig.~\ref{fig:steady-states}(a)), the RWA curve is always below the non-RWA one. The distance between them constantly increases as the non-RWA steady state for decoherence in the bath shifts up with both the bath frequency cutoff $\gamma_B$ and the bath coupling strength $\Delta_B$. The RWA curves exhibit saturation in both figures, but the non-RWA curves reach minimums, go up and approach the Markovian curves from below (more pronounced in Fig.~\ref{fig:steady-states}(b)). In the Markovian approximation the steady states seem insensitive to any changes of the bath spectral density parameters.

Now let us fix the bath parameters and change the stochastic field instead.  If we begin to gradually increase the stochastic field frequency cutoff from zero changing all the random processes simultaneously $\gamma_\nu=\gamma_F$ (Fig.~\ref{fig:steady-states}(e)), the field contribution in the resulting steady state starts growing and the steady excited state population of the TLS starts growing either, because the steady states for decoherence in the stochastic field always lie higher. At some value of $\gamma_F$ we reach the maximum, and after it the excited state population starts decreasing. The RWA applied to the interaction with the bath shifts the steady state down with respect to the one of the non-RWA curve, while the Markovian approximation gives much higher steady excited states populations for small frequency cutoffs and significantly overestimates the rate at which they decrease. The situation is similar if we change the field coupling strength in the same way $\Delta_\nu=\Delta_F$ (Fig.~\ref{fig:steady-states}(d)), but there is no maximum, and the curves continues to rise, approaching the Markovian curve from below.

The observed behavior can be explained via the magnitude of the environment spectral density (for the stochastic field it is the spectrum of corresponding random processes) in the vicinity of the TLS resonant frequency. The OU random processes and the Lorentzian bath have spectral densities with one peak. The OU process peak is located at $\omega=0$ and becomes wider and lower when the correspondent frequency cutoff increases. In Fig.~\ref{fig:steady-states}(d) we see how the stochastic field contribution in the steady states grows at first, causing the rise of the steady states curves, because the spectral density peak widens and the magnitude of the spectral density near the resonant frequency increases, then the contribution falls, because the process of the spectral density peak declining becomes dominating, and the curves go down too. The switch from the growth to the decline of the field spectral density near the resonance frequency determines the maximum of the RWA and the non-RWA curves.

In contrast, the peak of the bath spectral density is always located at resonance and widens when the bath frequency cutoff increases. In Fig.~\ref{fig:steady-states}(b) we see how the widening increases the magnitude of the bath spectral density in the vicinity of $\omega_0$ and the contribution of the bath in the steady states start growing, causing the curves to go down, because the TLS steady states for interaction only with the bath are located lower. Then the magnitude reaches its maximum and the contribution saturates. The subsequent rise of the non-RWA curves can be explained by the rise of the non-RWA curve for decoherence in the bath only (Fig.~\ref{fig:steady-states}(a)). 

Coupling strengths of both subenvironments affect only heights of the corresponding spectral density peaks, when a coupling strength grows, the peak grows either. It results in gradual increase of the magnitude of a spectral density near the TLS resonant frequency and can be seen in Figs.~\ref{fig:steady-states}(c) and \ref{fig:steady-states}(e). In Fig.~\ref{fig:steady-states}(c) the bath contribution increases, shifting the curves down, while in Fig.~\ref{fig:steady-states}(e) the field contribution increases, shifting the curves up. 

The situation becomes more complex if we stop using the restrictions $\gamma_\nu=\gamma_F$ with $\Delta_\nu=\Delta_F$ and allow arbitrary changes for each of the underlying random processes of the stochastic field. The detailed study of impact of each of the processes on the steady states will be presented elsewhere. 

\section{\label{sec:dm-evol}Density matrix evolution}
\begin{figure}[t]
\includegraphics[width=\linewidth]{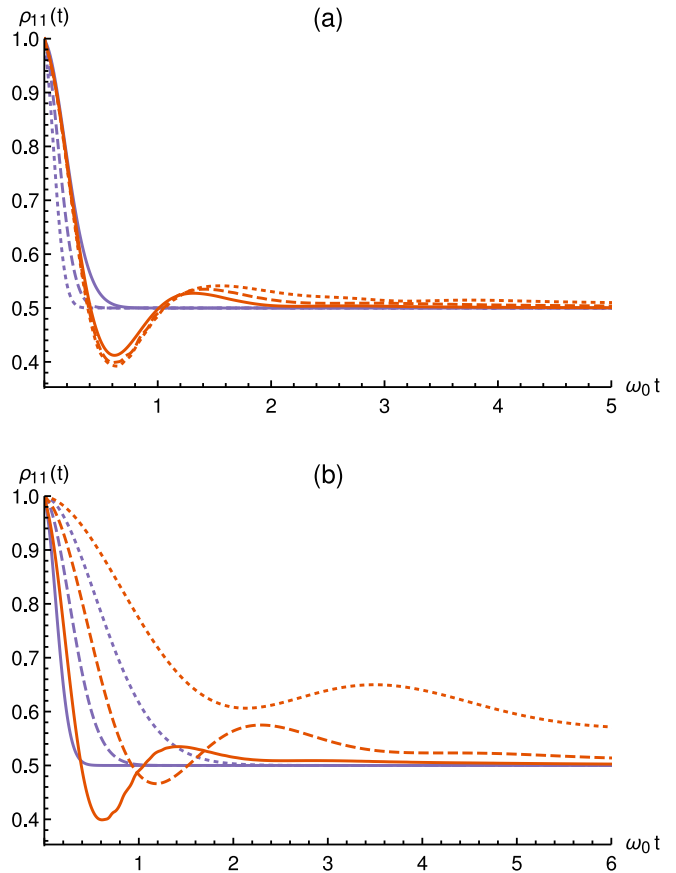}
\caption{
\label{fig:dm-evol-field}
Evolution of the TLS excited state population in the stochastic field in dependence on (a) the field frequency cutoff $\gamma_F$ and (b) the field coupling strength $\Delta_F$. Orange denote the non-Markovian curves, purple stands for the Markovian ones. In (a) $\gamma_\nu/\omega_0=\gamma_F/\omega_0=\{0.2, 0.4, 0.8\}$ and $\Delta_\nu/\omega_0^{1/2}=\Delta_F/\omega_0^{1/2}=1.6$, for \{dotted, dashed, solid\} curves respectively, and in (b) $\gamma_\nu/\omega_0=\gamma_F/\omega_0=0.4$, $\Delta_\nu/\omega_0^{1/2}=\Delta_F/\omega_0^{1/2}=\{0.4, 0.8, 1.6\}$. 
} 
\end{figure}

\begin{figure}[t]
\includegraphics[width=\linewidth]{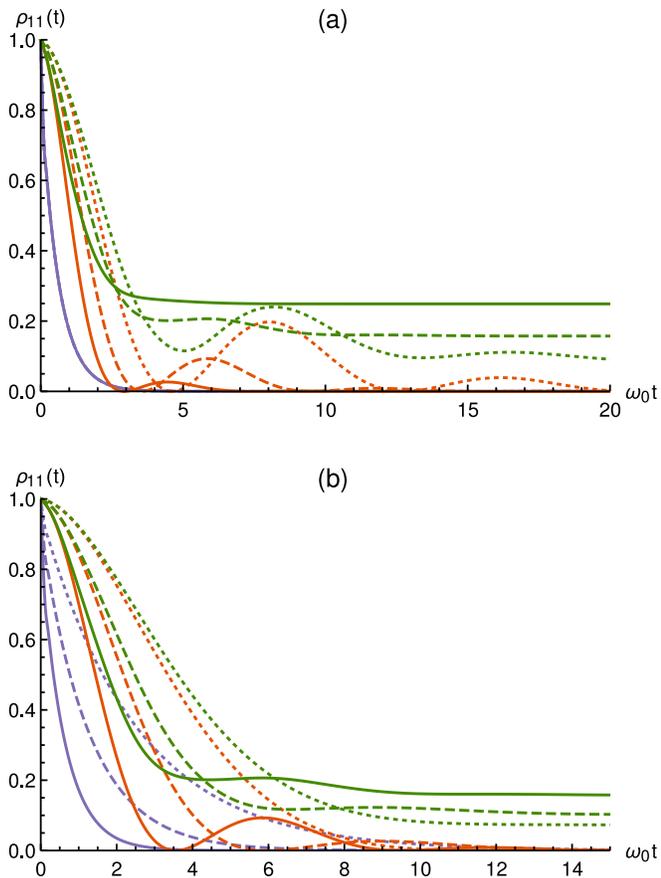}
\caption{
\label{fig:dm-evol-bath}
Evolution of the TLS excited state population in the Lorentzian bath (the stochastic field is off) in dependence on (a) the bath frequency cutoff $\gamma_B$ and (b) the bath coupling strength $\Delta_B$. Orange and green denote the RWA and the non-RWA couplings respectively, purple stands for the Markovian approximation. In (a) $\gamma_B/\omega_0=\{0.2, 0.4, 0.8\}$, $\Delta_B/\omega_0^{1/2}=1.6$, for \{dotted, dashed, solid\} curves respectively, and in (b) $\gamma_B/\omega_0=0.4\omega_0$, $\Delta_B/\omega_0^{1/2}=\{0.4, 0.8, 1.6\}$.
}
\end{figure}
\begin{figure}[t]
\includegraphics[width=\linewidth]{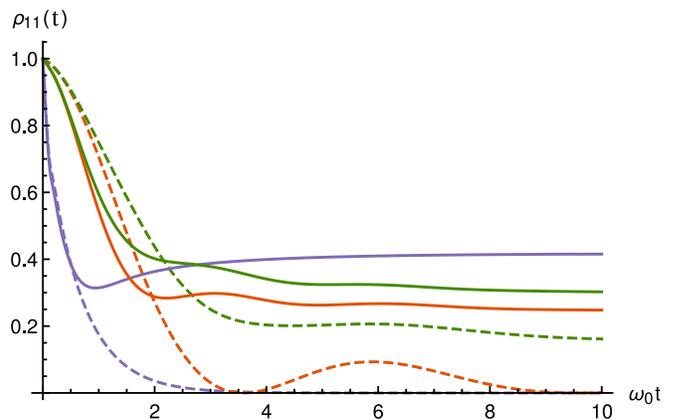}
\caption{
\label{fig:dm-evol-bath-field}
Evolution of the TLS excited state population for simultaneous interaction with the stochastic field and the Lorentzian bath (solid curves) in comparison with the case of interaction with the bath only (dashed curves). Orange and green denote the RWA and the non-RWA couplings respectively, purple stands for the Markovian approximation, $\gamma_\nu/\omega_0=\gamma_F/\omega_0=0.4$, $\Delta_\nu/\omega_0^{1/2}=\Delta_F/\omega_0^{1/2}=0.4$, and $\gamma_B/\omega_0=0.4$.
}
\end{figure}

An initially excited TLS placed in an equilibrium non-Markovian environment loses coherence due to the interaction with the environment, but the process is not monotone. At some point during the evolution the information backflow from the environment begins restoring the coherence, then the backflow weakens and the TLS starts losing coherence again. As a result, the oscillatory behavior emerges.

The evolution of the reduced density matrix from the factorized initial state (\ref{factorized-initial-conditions}) for different parameter regimes is presented in Figs.~\ref{fig:dm-evol-field} and \ref{fig:dm-evol-bath}. The curves corresponding to the non-Markovian evolution, both for the RWA and the non-RWA types of coupling with the bath, exhibit rapidly vanishing oscillations, which are more evident for the RWA curves, including all the non-Markovian curves for decoherence in the stochastic field (Fig.~\ref{fig:dm-evol-field}). The amplitude of the oscillations has a clear relation to shapes of the subenvironments spectral densities in the vicinity of the TLS resonance frequency $\omega_0$. If the environment spectral density is flat in the vicinity of $\omega_0$, which is the case of large frequency cutoffs, there are no oscillations. For example, the Markovian approximation curves in Figs.~\ref{fig:dm-evol-field}(a) and \ref{fig:dm-evol-bath}(a) exhibit no oscillations, because the Markovian approximation assumes that environment correlation times are small, which corresponds to large frequency cutoffs. If the environment spectral density is not flat in the vicinity of the TLS resonance, the oscillations appear. The dependence of the oscillations amplitude on the frequency cutoff value in Figs.~\ref{fig:dm-evol-field}(a) and \ref{fig:dm-evol-bath}(a) is more evident in case of decoherence in the bath, because the peak of its spectral density is located at the TLS resonance frequency. The coupling strength of the environment impacts the amplitude of the oscillations in the opposite way (Figs.~\ref{fig:dm-evol-field}(b) and \ref{fig:dm-evol-bath}(b)). 

The evolution becomes faster if the coupling strength increases, i.e. the minimums are located closer and the steady states are reached earlier (Fig.~\ref{fig:dm-evol-field}(b) and \ref{fig:dm-evol-bath}). The frequency cutoff impacts the speed of the evolution in a more complex way, there is a cutoff frequency for which the evolution speed is maximum (not shown).

The main difference between the RWA and the non-RWA curves in Fig.~\ref{fig:dm-evol-bath} resides in values of the minimums. The RWA curves have its minimums placed at the horizontal axis where also the stationary values are located, while the minimums of all the non-RWA curves are placed strictly above the corresponding stationary values. The stochastic field curves for sufficiently large couplings have their first minimum located below the stationary value (Fig.~\ref{fig:dm-evol-field}), which resembles the behavior of the non-RWA curves for decoherence in the bath. In overall, the non-RWA evolution of the reduced density matrix reminds the smoothed version of the RWA evolution.   

When the TLS interacts with both subenvironments simultaneously (Fig.~\ref{fig:dm-evol-bath-field}), the evolution becomes faster due to the presence of an additional decoherence channel, the first minimum is moved to the left and the stationary value is reached earlier. Because the steady state is shifted up by the stochastic field, the curves lie above the corresponding curves for decoherence in the bath only. Also the stochastic field significantly damps the oscillations, which is evident even for rather weak coupling strengths with the field in comparison with the coupling strength with the bath. The Markovian approximation shows the fastest decoherence among all the curves.  

\section{\label{sec:spectrum}Emission spectrum}
We obtain the equilibrium emission spectrums of the TLS by applying the Fourier transform to the two-time correlation function \(\langle\hat{\sigma}_+(t_2)\hat{\sigma}_-(t_1)\rangle\), where $t_2 > t_1$. The time $t_1$ is selected sufficiently big for the reduced density matrix evolution to reach its stationary phase, thereby the correlation function may be considered stationary. We calculate the stationary correlation function in the following way. First we propagate the initial state to the steady state, then apply the operator \(\hat{\sigma}_-\) to all density matrices \(\hat{\rho}^{(A)}_{\bm{m}}(t_1)\), lying in the TLS subspace, next the result is propagated to $t_2$, where \(\hat{\sigma}_+\) is applied. 

\begin{figure}[t]
\includegraphics[width=\linewidth]{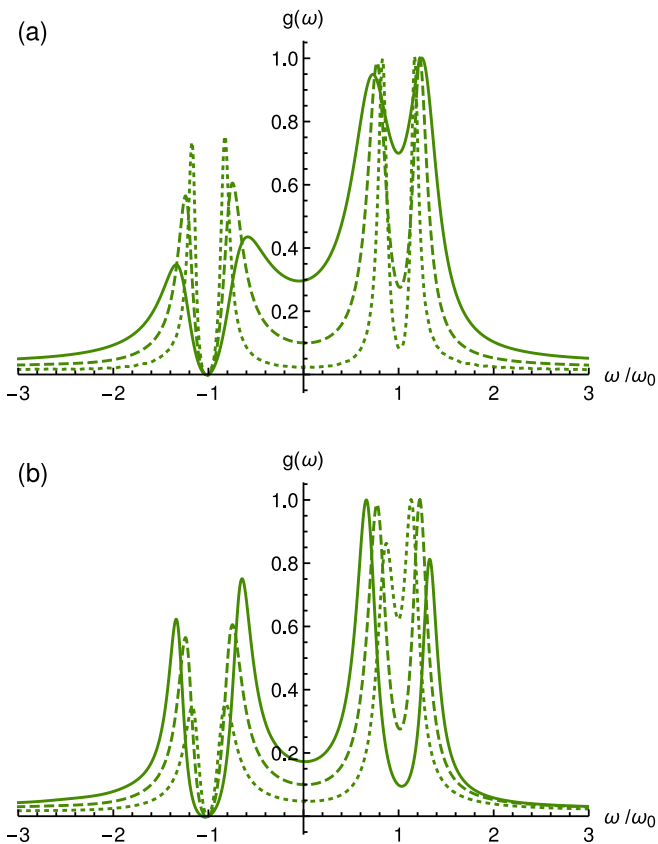}
\caption{
\label{fig:spectrum-bath}
Emission spectrums of the TLS in the Lorentzian bath, normalized by maximum values, in dependence on (a) the bath frequency cutoff $\gamma_B$ and (b) the bath coupling strength. In (a) $\gamma_B/\omega_0=\{0.1, 0.2, 0.4\}$, $\Delta_B/\omega_0^{1/2}=0.6$, for \{dotted, dashed, solid\} curves respectively, and in (b) $\gamma_B/\omega_0=0.2$, $\Delta_B/\omega_0^{1/2}=\{0.3, 0.6, 1.2\}$.
}
\end{figure}

The equilibrium emission spectrums for interaction with the Lorentzian bath can be obtained only in the case of the non-RWA coupling, because the steady states in the Markovian and the RWA approximations are completely unexcited and do not emit (Fig.~\ref{fig:spectrum-bath}). If the frequency cutoff is large, the non-RWA emission spectrum has one peak, which shifts to the right when the cutoff becomes smaller (not shown). At some cutoff value the top of the peak becomes a plateau and splits in two practically non-distinguishable peaks of non-equal intensity which are placed symmetrically with respect to \(\omega =\omega_0\) (not shown). Then the peaks move in the opposite directions, slightly declining, but stop at some value of the cutoff and start moving backwards while becoming more and more distinct (Fig.~\ref{fig:spectrum-bath}(a)). At the same time the overall intensity of the spectrum gradually decreases. As a result the two-peaked spectrums are fairly weak in comparison with the one-peaked spectrums. In Fig.~\ref{fig:spectrum-bath} we use the normalization by the maximum value, so the actual spectrum intensity is not shown. Potentially the appearance of the two peaks is explained by the presence of the two complex-conjugated coefficients \(\alpha^{(B)}_{k_1}\) and \(\alpha^{(B)}_{k_2}\) in (\ref{lorentz-sd-heom-coeffs}) instead of one, e.g. for decoherence in the stochastic field. Also there is a zero-intensity point for \(\omega=-\omega_0\) and two peaks on both sides of it. The peaks move towards each other when the frequency cutoff decreases. 

The dependence on the coupling strength with the bath is slightly more simple. If we increase it, the main peak widens and moves to the right, then splits in two peaks of unequal intensity (Fig.~\ref{fig:spectrum-bath}(b)). If we increase the coupling strength further, the right peak, which is the main peak, decreases, and the left (the side peak) grows, while moving in the opposite directions, the peaks resolution becomes better. For large coupling strengths the side peak becomes the dominant and approaches $\omega=0$.

\begin{figure}[t]
\includegraphics[width=\linewidth]{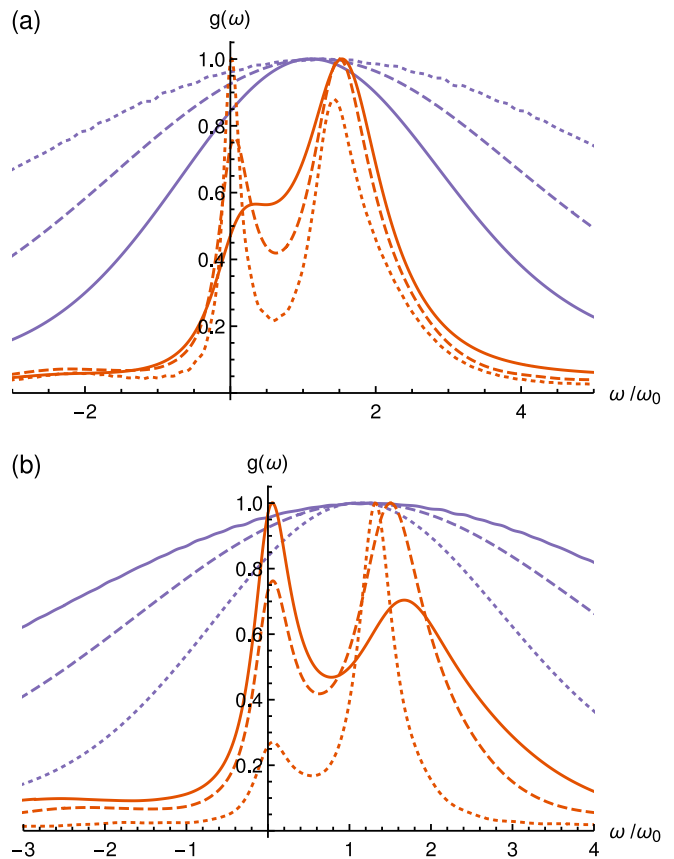}
\caption{
\label{fig:spectrum-field}
Emission spectrums of the TLS in the stochastic field, normalized by maximum values, in dependence on (a) the field frequency cutoff $\gamma_F$ and (b) the field coupling strength. Orange denote the non-Markovian curves, purple stands for the Markovian ones. In (a) $\gamma_F/\omega_0=\{0.1, 0.2, 0.4\}$, $\Delta_F/\omega_0^{1/2}=0.6$, for \{dotted, dashed, solid\} curves respectively, and in (b) $\gamma_F/\omega_0=0.2$, $\Delta_F/\omega_0^{1/2}=\{0.4, 0.6, 0.8\}$.
}
\end{figure}

For comparison, we show the equilibrium emission spectrums for decoherence in the stochastic ﬁeld in Fig.~\ref{fig:spectrum-field}. The spectrum differs qualitatively from the case of decoherence in the Lorentzian bath. The stochastic environment spectral density has a peak located at $\omega=0$ for any value of the frequency cutoff and the coupling strength. The peak becomes more distinct if the frequency cutoff lowers or the coupling strength rises. The resonance is clearly visible in Fig.~\ref{fig:spectrum-field}, where a side peak located at $\omega=0$ appears if the cutoff frequency is sufficiently small or the coupling strength is sufficiently large. At the same time, the main peak shifts to the right and widens. The spectrum energy redistributes from the main peak to the side peak, and the side peak grows while the main peak decreases.  

The Markovian approximation effectively considers the environmental spectral density flat (large frequency cutoffs), or, equivalently, it considers only a small region of the spectral density in the vicinity of the TLS resonance. If the spectral density resonance is located sufficiently far from the TLS resonance, the Markovian approximation loses the essential information about the peak existence. In Fig.~\ref{fig:spectrum-field} it results in wide one-peaked spectrums, which peaks are located in the vicinity of the TLS resonance frequency.

\begin{figure}[t]
\includegraphics[width=\linewidth]{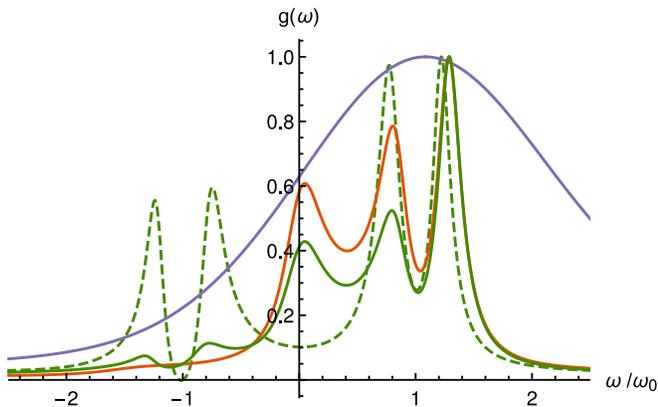}
\caption{
\label{fig:spectrum-bath-field}
Emission spectrums of a TLS for simultaneous interaction with the stochastic field and the Lorentzian bath (solid curves) in comparison with the case of interaction with the bath only (dashed curves), normalized by maximum values. Orange and green denote the RWA and the non-RWA couplings respectively, purple stands for the Markovian approximation, \(\gamma_{\nu }/\omega_0=\gamma_F/\omega_0=0.2\), \(\Delta_{\nu}/\omega_0^{1/2}=\Delta_F/\omega_0^{1/2}=0.2\), and $\gamma_B/\omega_0=0.2$, $\Delta_B/\omega_0^{1/2}=0.6$.
}
\end{figure}

In Fig.~\ref{fig:spectrum-bath-field} we show the impact of the stochastic field on the equilibrium emission spectrums for decoherence in the bath. The stochastic field causes the emergence of the zero-frequency peak, like it does in Fig.~\ref{fig:spectrum-field}, so the spectrum obtains the three-peaked form. Also it lowers the left peak of the doublet (the side peak in Fig.~\ref{fig:spectrum-bath}), widens it and shifts the main peak (the rightmost peak) to the right. The field smoothes the negative frequencies spectrum, making the zero point disappear, and increases intensity of the RWA curve so that it can be observed. The Markovian approximation is clearly inaccurate in the parameter regions selected. 

\section{\label{sec:conclusions}Conclusion}
We have studied non-Markovian evolution of a TLS in a composite environment consisting of two subenvironments, a zero-temperature bosonic bath characterized by the Lorentzian spectral density and a stochastic field of the Ornstein-Uhlenbeck type, and analyzed the impact of the rotating-wave approximation used for the interaction with the bath. 

It was shown that the steady states for decoherence in the bath depend on the coupling type used and the full interaction leads to different steady states in comparison with the cases when either the RWA or the Markovian approximation is used. We investigated the joint influence of the subenvironments on the steady states and found connections with the shape of the environment spectral density in the vicinity of the TLS resonant frequency.  

We demonstrated the dependence of the reduced density matrix evolution on the frequency cutoff and the coupling strength of the environment. In all cases, except the cases involving the Markovian approximation, the reduced density matrix exhibits the oscillatory behavior, the amplitude of which can be explained via shapes of the subenvironments spectral densities in the vicinity of the TLS resonance frequency. We showed that increasing the frequency cutoff smooths the oscillations and shifts the first minimum location to the right while increasing the coupling strength acts in the opposite way. Comparing the cases of the full and the RWA couplings to the bath, we found that the minimums of the oscillation in the RWA can be located below the stationary value in contrast to the case of the full interaction where this is not observed.

We investigated the dependece of the TLS equilibrium emission spectrums on the frequency cutoffs and the coupling strengths of the subenvironments and found that, if the TLS interacts only with the bath, the spectrums can have the doublet form in the positive frequencies domain and the doublet form for the negative frequencies domain at the same time, but the intensity of the spectrum is low. For interaction with both subenvironments, the spectrum can have three distinct peaks, one of which is located at the zero frequency, and the other two are located at the opposite sides of the TLS resonance.    

\bibliography{bibliography}

\end{document}